\documentclass[twocolumn,11pt]{article}
\usepackage{amssymb,amsmath,latexsym}
\usepackage[square,numbers]{natbib}
\usepackage{chemarr}
\usepackage{enumitem}
\usepackage{graphicx}
\usepackage{amssymb}
\usepackage{lineno}
\usepackage{amsmath}
\usepackage{sidecap}
\usepackage{wrapfig}
\usepackage{multirow}
\usepackage{float}
\usepackage{cuted}
\newcommand{\MSun}{{M_\odot}}

\newcommand{\DD}{\frac}
\newcommand{\beq}{\begin{equation}}
\newcommand{\eeq}{\end{equation}}
\newcommand{\ben}{\begin{enumerate}}
\newcommand{\een}{\end{enumerate}}
\newcommand{\bit}{\begin{itemize}}
\newcommand{\eit}{\end{itemize}}
\newcommand{\barr}{\begin{array}}
\newcommand{\earr}{\end{array}}
\newcommand{\bc}{\begin{center}}
\newcommand{\ec}{\end{center}}
\newcommand{\Msun}{{\,{\cal M}_{\odot}}}

\begin{document}

\twocolumn[
  \begin{@twocolumnfalse}
  \title{Does our  universe conform with the existence of a universal maximum energy-density  $\rho^{uni}_{max}$ ?}

\author{\thanks{E-mail:AHujeirat@uni-hd.de}~Hujeirat  A.A.  \\
IWR, Universit\"at Heidelberg, 69120 Heidelberg, Germany \\
 }
\maketitle
\begin{abstract}
\noindent Recent astronomical observations of high redshift quasars, dark matter-dominated galaxies, mergers of
neutron stars, glitch phenomena in pulsars,  cosmic microwave background and experimental data from hadronic colliders do not rule out, but
they even  support  the hypothesis that the energy-density in our universe most likely is upper-limited by $\rho^{uni}_{max},$  which is predicted to lie between
$2$ to $3$ the nuclear density  $\rho_0.$ \\
\hspace{0.5cm}
Quantum fluids in the cores of massive NSs with $\rho \approx \rho^{uni}_{max}$ reach the maximum compressibility state, where they become insensitive to further compression by the embedding spacetime and undergo a phase transition into the purely incompressible gluon-quark superfluid state.
A direct correspondence between the positive energy stored in the embedding spacetime and  the degree of compressibility and superfluidity of the trapped matter is proposed.\\
In this paper relevant observation signatures  that support the maximum density hypothesis are reviewed, a  possible origin of   $\rho^{uni}_{max}$ is proposed  and finally  the consequences of this scenario on the spacetime's topology of the universe as well as on the mechanisms  underlying the
growth rate and power of the high redshift QSOs are discussed.


\end{abstract}
 \end{@twocolumnfalse}
 ]
\vspace{1.9 cm}
\section{Introduction}
Very recently we have witnessed  a historical breakthrough in observational astronomy: the beginning of the Multi-Messenger  era. The
accurate and successful calibration of instrumentations operating remotely at different frequency regimes has successfully enabled us to detect
astronomical events with unprecedented accuracy. Thanks to modern communication technologies that enable pairing remote telescopes and
form global networks of virtual telescopes that provided the highest resolution imaging currently possible at any wavelength
in astronomy \cite[see][and the references therein]{Abbott_MMessenger2017,Falke_MMessenger_VLB2012}. Based thereon, observation of highly energetic astrophysical events, such  as mergers of neutron stars and black holes,
GRBs, Kilonova resulting from NS-mergers,  imaging the event horizons of black holes   \cite[see][and the references therein]{Abbott_MMessenger2017,Nichol_Kilonova2021,Psaltis_EHT2020} as well as quasars at extremely high redshift \cite{Nichol_Kilonova2021}
 delivered a lot of data that enabled us to gain deeper insight of the underlaying physics. Other breakthroughs have been recorded on the micro-scales,
 such as the Higgs bosons at the Large Hadronic  collider (LHC), and the fluidity character of quark-gluon plasmas at the RHIC
  \cite[see][and the references therein]{McInnes_QGP_ReynoldsNumber2016,Pasechnik_QGPTransition2017,Shuryak_QGPTransition2004}. \\
 On the other hand, while  modern observations are continuously shedding light on the huge diversity of the state of matter both on the micro and macro-scales,  they also raise new questions related to the validity of solutions of problems that have been thought to have settled. \\
Based on the recent theoretical and observational studies, it is argued in this paper that:

 \begin{figure}[t]
\centering {\hspace*{-0.15cm}
\includegraphics*[angle=-0, width=6.0cm]{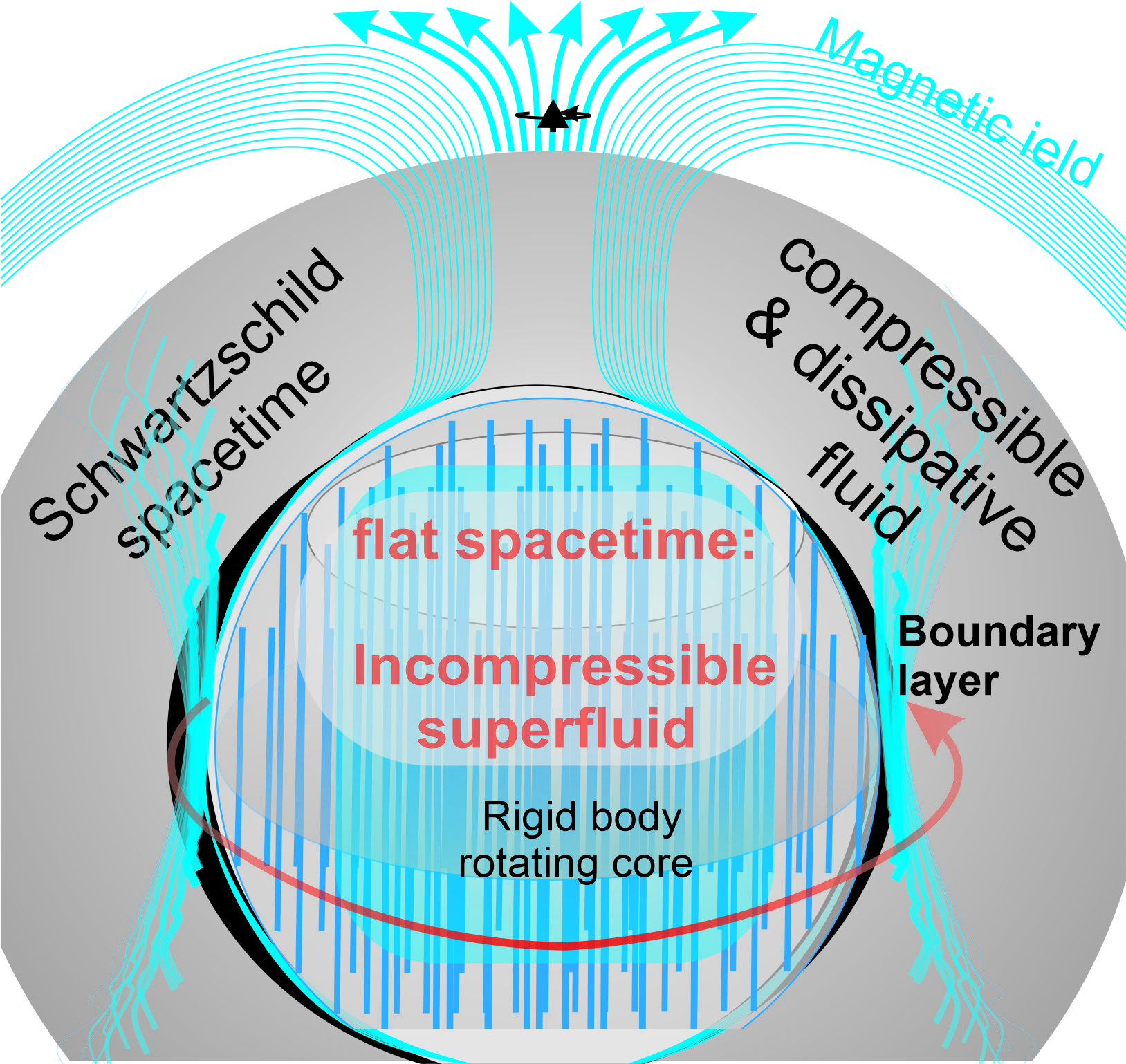}
}
\caption{\small A schematic description of the internal structure of a neutron star and the topology of the embedding spacetime.
There are three different regions that are relevant for the present discussion: the central core, which is made of incompressible superfluid and 
imbedded in a Minkowski-type spacetime. The surrounding shell, inside which the matter is compressible, dissipative and embedded in
a Schwarzschild-type spacetime. The geometrically thin boundary layer between the core and the surrounding shell, where  matter and  spacetime 
repeatedly undergo phase transitions.
}\label{PulsarBimetric}
\end{figure}
\begin{itemize}
  \item If  $\rho^{uni}_{max}$ indeed exist, then the remnant of the merger of the two neutron stars in  GW170817 should be a massive neutron stars (NS) whose core is made of an incompressible quark-gluon superfluid
           that obeys the laws of quantum field theories \cite[see Fig. 1 and reference:][]{HujSam2020GW170817}.    However, the ambient media remain compressible and dissipative, though these properties are doomed to disappear at the end of the luminous lifetime of NSs, rendering them  invisible.\\
           These  cores appear to be generic for almost all massive NSs, whose masses and dimensions are set to grow as NSs age. Within the context of general
           relativity, a physically consistent  and causality-preserving treatment of the two totally distinct fluids:  incompressible-superfluid and compressible-dissipative normal fluid, is possible,  if the embedding  spacetime is of a bimetric type: a Minkowski flat at the background of
            the core surrounded by a Schwarzschild curved one.
  \item   The above argument is extended to show that the history and large-scale structures of our universe may indeed tolerate
  the existence of $\rho^{uni}_{max}$.  Also,  I address and discuss several relevant observational signatures both on the micro and macroscopic scales  that  support the maximum energy density  hypothesis.
  \item  If such a density upper-limit does exist, then what could be the underlying physical mechanisms that may lead to the existence of $\rho^{uni}_{max}$?
\end{itemize}
\section{Observational signatures }
\subsection{The origin of power of high redshift QSOs  and their growth problem}
Early large-scale structures are considered to have formed during the first two billion years after the big bang.
Indeed, the three UV-emission lines from the luminous star-forming galaxy GN-z11 which has been recently analysed
using Hubble Space Telescope (HST) imaging data,  most likely correspond to emission at z = 10.957+/-0.001, i.e. to roughly 420 Myr after the big bang  \cite[see][and the references therein]{Fan_QSO_Redshift2019,Oesch_Z11Galaxy2016}.
Taking into account that GN-z11 is roughly 100 less massive than our Galaxy, then GN-z11 is expected to  host a SMBH
at its center.
Similarly, recent observations of   the quasar ULAS J1342+0928 reveal that its redshift  is $z = 7.54 $ \cite[see][and the references therein]{Yang_SMBH_HR7.5_2016}, which implies
that  the object must have been there already when the universe was 700 Myr old. Such rapid growths cannot be explained, unless the seeds are
massive BHs  of at least $10^6 \MSun$ \cite[see][and the references therein]{Shimasaku_SMBH_growth_HRshift2016,Bromm_SMBHd2017} . Therefore primordial black holes that were formed either within the first $10^{-23}$ s
or  during the first $10^{-5}$ s after the big bang should be ruled out, due to lacking observational evidence for their existences
\cite[see][and the references therein]{Carr2010}.\\
Irrespective of the underlying growths mechanisms,  their mass increase should follow the equation of mass conservation:
 \beq
 \DD{d \mathcal{M}}{dt} = \sum  \int_S\bar{\textbf{ f}}\cdot d\textbf{s} = a_0 (\DD{\beta_v}{\beta^4_c})\rho \mathcal{M}^2 +
\dot{\mathcal{ M}}_{ext},
 \eeq
 where $a_0 = 16\pi G^2/c^3 \approx10^{-6}$ and $ \beta_v ,  \beta_c, \rho $ correspond to the relative fluid velocity, $v/c$,
 sound speed, $V_s/c$ and the density of the inflowing matter through  the boundary in units of $10^{-10}$ g/cc, respectively.
 Here $\mathcal{ M}/t$  are in units of $5\times 10^7\, \MSun$ per year. $ \dot{\mathcal{ M}}_{ext}$ denotes other growth mechanisms, such as direct collapse of primordial gas clouds, runaway collisions of dense star in clusters,  hyper-Eddington accretion as well as episodic galaxy mergers
 \cite[see][and the references therein]{Bromm_SMBHd2017}.
 Assuming $d({\beta_v \rho/\beta^4_c})/ d \mathcal{M} <<1,$ then the time-dependent mass should evolve as:
 \beq
 \DD{\mathcal{M}(\tau)}{\mathcal{M}_0} =  \DD{1}{ 1 -  a_0 (\DD{\beta_v}{\beta^4_c})\rho\,  \tau\, \mathcal{M}_0 },
 \eeq
 where $ \tau$ corresponds to the elapsed   time for a SMBH of  initial mass $ \mathcal{M}_0$  to reach
$\mathcal{M}(\tau).$ For  $\mathcal{M}(\tau)/ \mathcal{M}_0 = 100$ and $\mathcal{M}_0=1,$
we obtain roughly:
\beq
\tau_{100} \approx \DD{10^6}{\rho} \times  (\DD{\beta^4_c}{\beta_v}) =
 \left\{
  \barr{rl}
   > 10^9 &     R_A = R_{LSO} \\
           &     ({\beta_v}={\beta_c} \approx 1) \\
   \mathcal{O }(10^9) & \textrm{\small Super-Eddington Accr.} \\
               &  \textrm{\small collapse of clouds,} \\
               &  \textrm{\small collision and merger}.
  \earr
 \right.
\eeq
$R_{LSO}$ here is the last stable orbit, where ${\beta_v}={\beta_c} \approx 1.$  The  Eddington accretion is used to
estimate the density $\rho$. In the case of super-Eddington accretion caused by direct
collapse of clouds, star collision and mergers, the matter is expected to be increasingly hotter as the center of mass is approached, hence more
difficult to compress, thereby giving rise to  slow and episodic accretion.\\
 The conclusion here is that current scenarios are incapable of providing viable  solution to the rapid growth of SMBHs within the
 first 600 Myr after the big bang. This problem will soon become more prominent, when advanced observations shed the light on the existence of
 QSOs at $z \geq 10,$ which would logically  imply that their existence of QSOs at $z \gg 10$ cannot be ruled out. \\
 Moreover,   high redshift QSOs may have been there already before the big bang, and that their nature must be more complicated than being objects with  mathematical singularities  at their centers.\\

\subsection{Compact objects and the mass-gap }
Astronomical observations reveal a gap in the mass spectrum of compact objects:
neither black holes nor NSs have ever been observed in the mass range $[2.5\MSun \leq \mathcal{M}_{NS} \leq 5\MSun].$
 Even the nature of the remnant of the well-studied NS-merger event, GW170817, whose mass is predicted to be around
 $2.79\,\MSun,$ is still not conclusively determined on whether it is a massive NS or a stellar-mass BH
 \cite[see][and the references therein]{HujSam2020GW170817,AbbottNSProberties2019}. \\
 Recently it was suggested that when  massive NSs cooldown during their cosmic evolution, their incompressible superfluid cores should grow both in mass  and volume to finally fade away as dark energy objects  \citep[DEOs; see][for further details]{Hujeiratmetamorph2018,HujSam2020B,HujSam2020CrabVela}. Based on observational data of glitching pulsars, and specifically
 of the Crab and Vela pulsars, this scenario appears to successfully explain various aspects of the glitch phenomena, such as the growth of mass and volume of their cores, as well as of the mechanisms underlying the under- and overshootings observed to accompany their prompt spin-up
 \cite[see][and the references therein]{HujSam2020CrabVela,Ashton_VelaOvershooting2019}.
 Recalling that  the Crab and Vela  are considered to the most well-studied pulsars in our galaxy, the  remarkable agreement of the scenario with observations was possible only, because of the  assumption that their cores are made of incompressible quark-gluon superfluids.\\
 Based thereon, isolated massive NSs either in the local  or  early universe should metamorphose into DEOs, whose embedding spacetimes is of Minkowski type.
\subsection{Dark matter in the early universe}
Dark matter (DM) in cosmology  is an essential pillar for understanding the entire cosmic web: starting from structure formation, clusters of galaxies
and galaxy formation down to stellar-mass objects. Except for gravitational interaction with normal matter,   the physics of dark matter continues to be a conundrum in understanding our universe.\\
For the present discussion, we select three properties that support the constant density hypothesis:\\
When a massive ultra-compact object\footnote{A family of objects that include pulsars, neutron stars, and magnetars.} (UCO)  run out of all their secondary energiesc\footnote{All other types of energies except the rest energy},
  the embedding spacetime should have compressed the matter inside their interiors up to the maximum limit, at which matter undergo
  a phase transition into incompressible gluon-quark superfluid, which corresponds to   the lowest possible quantum energy state. Objects
  of this type are both electromagnetically and gravitationally passive and are classified here as  DEOs.\\

  Indeed, the ultra diffuse galaxy, Dragonfly 44, is considered to be a  dark matter dominated galaxy \cite[see][and the references therein]{Arbey_DM2018,Bogdan_D44_2020,Argueelles_SMBH_In_DM_2021}.
  Its dwarf-like shape and the extremely low surface brightness indicate that it might be the relic of an old massive galaxy,
  in which the stars have run out of their secondary energies during the course  of their cosmic evolution \cite{Bogdan_D44_2020}. Also,
  it was argued that the lack of X-ray emission is an  indication  that most of the compact objects  in the galaxy
  are relatively old and therefore turned dark by now. Assuming star formation rates and supernovae statistics to apply
   for this galaxy during its active phase in the past, then roughly  $1\%$  of its population must be in the form of NSs and $0.1\%$
    stellar black holes. While BHs need to swallow matter in order to be detected, NSs may still
  eject highly relativistic particles from their polar caps and therefore  at least $0.001\%$ of the NS population
   should be detectable.   However, such signatures appear to be missing in the D44 galaxy.\\
  In the context of our scenario,  the old and massive NSs in D44 must have metamorphosed into DEOs that subsequently migrated
  either outwards into the halo or inwards where they conglomerate to form a cluster of
  massive DEOs. The time duration required
  for these objects to  randomly migrate from central region of a galaxy to the surrounding halo or vise versa may in random manner is of order
   $ \tau_{diff} \sim L^2/\nu_{eff},$ where $L$ is the average radius of the galaxy and
     $\nu_{eff}$ is a measure of their  gravitational scattering, $ \sigma_{DNS-Star},$ times the duration
     of their momentum exchange with other objects  $\tau_{DNS-Star}.$ Assuming the masses of both types of objects
     to be of order the solar mass, then $ \sigma_{DNS-Star} \approx 10^{-4}R^2_\odot.$ On the other hand,
     $\tau_{DNS-Star}$ can be set to be of order $ \sqrt{\sigma_{DNS-Star}}/V_{rel},$ where $V_{rel}$ is the relative velocity
     of the two arbitrary encountering objects\cite[see][for further details]{Gustafsson_GravScattering2016}. Setting a maximum relative velocity of $100 $ km/s, we obtain
     $ \tau_{diff} $ of order $10^3$ times the age of the universe.\\
     Consequently, massive NSs that run out of their secondary energies and became DEOs  either prior to or  post the big bang
     may be considered as DM-candidates. Their poor collisions and mergers  may be attributed to the  topology of  the bimetric spacetimes embedding
     DEOs. This picture is in line with recent observational and theoretical studies indicating that DM may have been
present already in the pre-big bang era and/or  during cosmic inflation \cite[see][and the references therein]{Barkana2018,Tenkanen_DM2019}.  However, unlike our present  macroscopic approach,
the underlying assumption of the former scenario is that DM should be made of microscopic particles  of unknown origin
that emerged out of  scalar fields.
\subsection{Merger of NSs and  glitching pulsars}
Observations reveal that NSs occupy a relatively narrow mass range that lies between $1.2\MSun$ and $2.2\MSun,$ though the error bars in many cases are relatively large. Despite the use of modern instrumentations and global coordination of multi-messenger observations, their radii,
which should decrease with increasing mass, are still rather uncertain, .\\

\begin{figure}[t]
\centering {\hspace*{-0.15cm}
\includegraphics*[angle=-0, width=5.25cm]{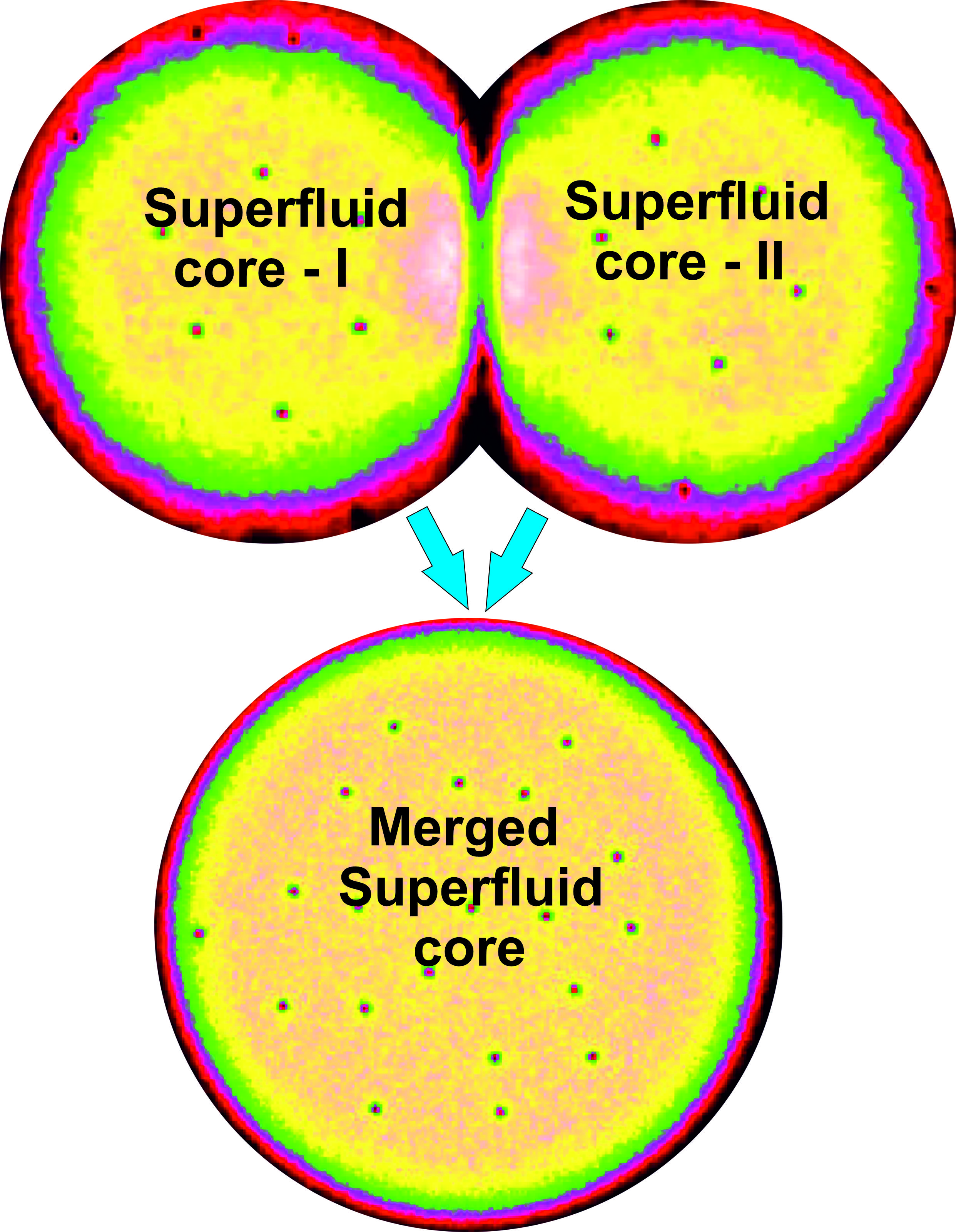}
}
\caption{\small By means of numerical solving the modified time-dependent Gross-Pitaevskii equation in 2D, the merger
of two superfluid cores trapped in a symmetric external potential have been calculated. The energy loss due to merger of the two cores
was found to decrease with decreasing the numerical diffusion, i.e. with increasing the spatial and temporal accuracies.
}\label{GPE-SFluidCores}
\end{figure}

Both theory and numerical calculations indicate that the central density of  neutron stars must be larger than the nuclear $\rho_0.$
However  probing  the state of ultracold supranuclear dense matter under terrestrial conditions is unfeasible and  the governing physics
 remains rather uncertain. Therefore, when modeling the internal structure of NSs, the regularity condition usually imposed at the centers of UCO is actually identical to  the incompressibility condition,  which is a limiting case of:
\beq
        \DD{\Delta \varepsilon}{\Delta r}  \xrightarrow[\Delta r \rightarrow 0]{} \nabla \varepsilon =0.
\eeq
Microscopically $\Delta r$ is limited from below by  the average separation between two arbitrary baryons  $\Delta_{bb}.$
Using  $\Delta r = \Delta_{bb} $ and $ \Delta \varepsilon \Delta t \approx \hbar, $ we obtain that $\Delta r \gg 10^{10}\times \Delta_{bb} .$
Among others, this implies that incompressibility is a macroscopic property, but that quantum fluids must fulfill on the microscopic scale
 at $r=0.$
In this case, the pressure here ceases to behave as a local thermodynamical quantity and it becomes solely a Lagrangian multiplier: a mathematical
term that dictates the global dynamics of the quantum fluid, which in turn depends strongly on the topology of the embedding spacetime. This pseudo-pressure  should be carefully constructed, so that its transition  into the ultrabaric regime remains forbidden,  $dp/as d\varepsilon \geq 1,$
and therefore causality condition is grossly violated \cite[see Fig. 7 in ][]{Hujeiratmetamorph2018}. \\
Indeed, we consider the  following observational signatures to support our scenario:
\bit
\item Isolated NSs with $\mathcal{M}_{NS} \geq 2.5\,\Msun$  are practically missing and theoretically difficult to  model, irrespective
         of  EOSs used.
\item Based on the multi-messenger observations of the NS-merger event GW170817, the remnant should have a mass of about
        $2.69\, \MSun,$ though its nature is still an open question.  In particular, observations did not conclusively rule out that
        the remnant might be a massive  NS. In fact recent numerical calculations of merging two incompressible superfluid cores
        trapped in a potential that mimics an embedding curved spacetime show that such cores are capable of merging without
        decaying \cite[Fig.\ref{GPE-SFluidCores}; see][]{Sivakumer_SFluidCoreaMerger2021}. Combining this numerical observation with the possibility that the merger of the overlying compressible and dissipative ambient media of both objects may  switch off the dynamo action of the remnant, thereby diminishing most magnetic activity not a short period.

\begin{figure}[t]
\centering {\hspace*{-0.15cm}
\includegraphics*[angle=-0, width=6.15cm]{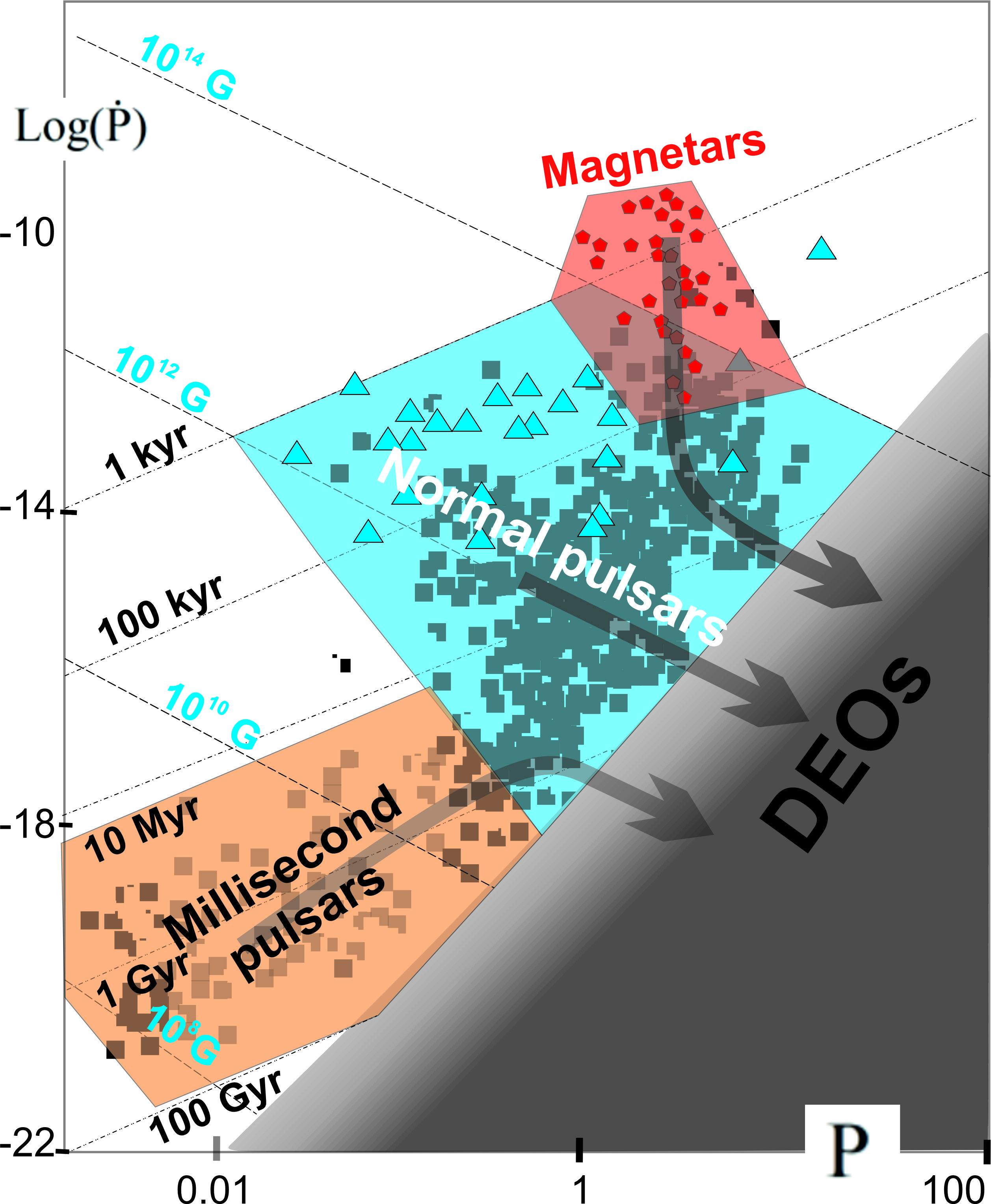}
}
\caption{\small  The derivative of the period time, $\dot{P},$ versus period ${P}$ of the population of  UCOs and
their tendency to migrate (marked with arrows) toward the lower-right region, where they end up as DEOs. There are three distinct classes that can be identified: millisecond pulsars (gray squares/orange region), glitching and normal pulsars (triangles \& squares/blue region) and magnitars
 (pentagons/red region). All types of UCOs should migrate into the lower right region, where they end up as DEOs. While  migration of strongly magnetized isolated and massive UCOs should proceed relatively fast, those with intermediate  masses in binaries and/or in  accretion mode generally need much longer time to migrate.
}\label{PulsarPopulation}
\end{figure}

\item The discrete events of glitches observed in pulsars are considered to be the responses of their cores to the long-term cooling of the objects.
       Moreover, the majority of    glitching pulsars appear to have  the following common features:
        \bit
         \item Glitches are associated with  the evolution of pulsars and young NSs, they occur instantly but the corresponding
                       relaxation time increases as they age. Recently,  accurate observations revealed certain association of under and overshootings
                       with the prompt spin-up events of glitching  pulsars \cite[see][and the references therein]{HujSam2020CrabVela,Ashton_VelaOvershooting2019}.\\
         \item  In the $\dot{P}-P$ diagram, almost all observed glitching pulsars appear to be younger than $10^7$ yr and
                        their magnetic fields are stronger than $10^{11}$ Gauss.
         \item The reoccurrence of glitches  decreases with pulsar's age \cite{Roy_GlitchActivityA2012,Espinoza_GlitchActivityB2011}.
         \item  UCOs, with  magnetic fields larger than the quantum critical value $B > B_c = m^2 c^3/e\hbar \approx 4.4\times 10^{13}$ G,
                  are found to occupy a narrow range in $B-$age diagram,  where the majority of magnetars are found to be younger than $10^5$ yr
                  \cite[Fig.\ref{PulsarPopulation}; ][and the references therein]{Gotthelf_PulsarPopulation2013,Ashton_PulsarPopulation2017,Ng_PulsarPopulation2015}.
          \eit
\eit

As it was shown in \cite{HujSam2020GW170817,Hujeiratmetamorph2018,Huj2020ConMap}, these features may easily be  explained  by invoking the bimetric spacetime scenario. Accordingly,  cores
of massive NSs are made of incompressible superfluids embedded in Minkowski spacetimes, whereas the surrounding
compressible and dissipative media  is embedded in Schwarzschild-type spacetimes. This scenario is consistent with general
relativity as the mass-energy inside the cores dictates the topology of the embedding spacetime and vise versa.\\

\subsection{The missing first generation of massive neutron stars}
Within the framework of  the  big bang scenario, the first generation of stars must have formed approximately
500 Myr after the big bang \cite[see][and the references therein]{Bromm_FirstStars2013}. The progenitor clouds were made mainly of atomic hydrogen with an average
temperature of about $1000$ K. This relatively high temperature enables cloud  to collapse under their own
self-gravity, if the  mass content is  larger than the critical Jeans mass amounting to $10^2$ up to $10^4$ solar masses.
Hence the resulting stars must have been extraordinary luminous and therefore short-living, and expected to subsequently
collapse into BHs or massive NSs. While BHs are generally accepted to be the remnants of collapsed  population III stars,  NSs
however have been neither investigated nor observed. Moreover, given the relatively large dimensions and long time
scales characterizing the progenitor clouds in combination with rotation, it is not at clear, why the centrally collapsing portion of the cloud with its relatively short dynamical time scale should proceed self-similar and fail to blow away  the outer slowly contracting shells or forbids fragmentation
of the collapsing cloud.  We note that the effect of numerical diffusion
in the simulations of the collapsing clouds is to enhance coupling of the different parts of the cloud and facilitate their direct collapse.
In real life, however, the cores of collapsing clouds will be much hotter and the advected magnetic fields via ambipolar diffusion will
soon be strong enough to fragment the cloud. Such sophisticated and robust numerical solvers are still to be developed, as important
mechanisms, such as radiation transfer, ambipolar diffusion, multi-component fluids, large scale magnetic fields, implicit time-stepping
that respect causality are still not accounted for.\\
 Nevertheless, to form  a BH, the spacetime embedding  the progenitor should have compressed
the  central core and  increased its central density up to the critical density $\rho_{cr},$ beyond which normal repulsive forces
fail to oppose further compression. In this case, the event horizon $r_H$ should surpass the actual radius of the central core,  and the enclosed average density must surpass the critical value:
\beq
\rho_{cr} \approx 1.6 \times 10^{27} \DD{1}{r^2_H}.
 \label{BHDensity}
\eeq
For $r_H$  of order $10^6$ cm,  one obtains $\rho_{cr}\approx 6~ \rho_0, $ where $ \rho_0$ is the nuclear density.\\
Obviously,  the reaction time scale of the core at the verge of collapse is of order  micro-seconds, which is comparable
to the light crossing time. Hence the core is well-equipped to oppose all types of external forces including compressional forces.
In this case, the accumulated matter on the surface most likely would bounce  rather than triggering a collapse of  the object
into  a BH. Indeed, most sophisticated numerical modeling of core-collapse supernovea  give rise to bouncing rather than
to direct collapse into a stellar BH \cite[see][and the references therein]{Fischer_BounceSN2021}, provided the domain of calculations does not contain an apparent horizon, which is
manually created by changing the EOSs long before $\rho^{uni}_{max}$ has bee reached  and/or surpassing  the maximum compressibility limit \cite[see][and the references therein]{Chon_ApparentHorizon2021}.  Moreover, most of today numerical solvers  are still neither capable of modeling incompressible supranuclear dense matter inside the cores of massive objects that are on the verge of collapsing into BHs nor treating the topology of the
embedding spacetime in a self-consistent manner  and therefore are irrelevant for the present discussion.\\
On the other hand, the multi-messenger  observations of NS merger event,  GW170817,  suggest
that violent merger of NSs may not necessarily lead to BH formation,  but that forming a dynamically stable massive NS
with $ \mathcal{M}_{NS}> 2.79\, \MSun$ should not be ruled out either \cite{HujSam2020GW170817}. The luminous lifetime
of such a massive NS would be extremely short  and would soon turn invisible DEO.\\
Consequently, as the first generation of stars must have been massive, their collapse should have generated a numerous number of massive NSs.
Their luminous lifetime would  be relatively short and therefore they should have entered their final evolutionary phase as DEOs long ago.
   On the other hand,  supernovae  statistics predict that  at least  $ 1\%$ of star populations in star-forming clouds should be NSs. Although this rate is expected to be even higher in the early universe, yet isolated NSs older than one Gyr have not been observed. The topology of the spacetime
 embedding DEOs is of a bimetric type, which in turn prohibits their merger and makes them to excellent  BHs and DM-candidates.

\subsection{Hadronic collisions: LHC and RHIC}
Among many recent explorations related to particle collisions in labs,  there are several discoveries that should be relevant for the present discussion:
\begin{itemize}
  \item Fluidity of quark-gluon plasmas\\
  Quark-gluon-plasmas  emerging from  smashed nuclei both at the Relativistic Heavy Ion Collider (RHIC) and the Large Hadron Collider (LHC)
  were verified to  behave nearly as perfect liquids \cite[see][and the references therein]{Pasechnik_QGPTransition2017,Shuryak_QGPTransition2004}.
 This was a turning point in physics, as the entropy of such colliding particles under normal  laboratory conditions is extremely high
and  the outcome is expected to be a dilute gas: an assembly of loosely connected particles  that running out of the center of collisions
 almost in spherical symmetric patterns. \\
Recalling that QGPs are imperfect due  to unitarity,  the cross-sections of colliding particles ought to be bounded from below.
Indeed, based on infinitely coupled supersymmetric Yang-Mills gouge theory (SYM) and using anti-de Sitter space/conformal field theory (ADS/CFD) duality conjecture, the lower-limit was verified to be:
\beq
(\DD{\eta}{s})_{SYM} = \DD{1}{4\pi},
\eeq
which is roughly equal to the lower-limit obtained using the uncertainty principle \cite{Mueller_QGP_LiquidUncertainty2007}.\\
While the physics governing particle collisions under terrestrial conditions are totally different from those
at the center of massive NSs, the emergence of quarks and plasma out of nucleons as liquids can be considered
as a strong evidence for the here-proposed hypothesis. Indeed, let the velocity of particles  immediately after
collisions be of order the sound speed: $V \approx V_s =c/\sqrt{3}.$ This yields a dynamical time
scale $ \tau_d = \lambda/V_s= \alpha_\lambda r_b/c .$ On the other hand, the viscous time scales as:
\beq
\tau_{vis}\approx \DD{\lambda^2_\lambda}{\nu_{vis}} =  \DD{\alpha^2_\lambda r^2_b}{\alpha_{vis} V_s <\ell_{vis}>} =
  \DD{\alpha^2_\lambda }{\alpha_{vis} \alpha_{\ell}} \tau_d,
\eeq
where $\alpha_{vis}$ denotes the effective viscous velocity in units of the sound speed and $ \alpha_{\ell}$ the corresponding
viscous length scale in units of $r_b.$ Given that elliptic flows start shaping already during the initial expansion, then
$  \DD{\alpha^2_\lambda }{\alpha_{vis} \alpha_{\ell}}$ should be of order unity, which yields the Reynolds
number:
\beq
Re = \DD{\tau_{diff}}{\tau_{dyn}} \approx 1.
\eeq
Going back in time, this result would  imply that the confined  QGPs, inside the sample of  nucleons under lab conditions,
 are dissipative liquids with non-vanishing viscosity over entropy, i.e. $0<\eta/s<1.$\\
 Therefore  the energy state of the QGPs here should  decay into a lower energy state to finally hardonize into stable baryons.\\
 However, in low Reynolds number flows, the effect of viscosity is to smooth out irregularities and prohibits the
 formation of distinct flow patterns, such as the micro-elliptical flows of deconfined QGPs, whose patterns
  considerably deviate from a pure spherical expansion. This implies that the dynamical time scale, on which particle collisions at the LHC/RHIC  occur, are still too long compared to the  time scales on which gluon effects are operating during the hardonization of the QGP.\\
 In the case of  NSs, as they cooldown on the cosmic time, their heat contents decrease and the supranuclear dense matter
 inside their cores approach the limit of maximum compressibility. Here both the temperature and entropy should go to the zero limit.
From statistical thermodynamical point of view,  the number of possible microstates of  constituents of incompressible superfluid
reduces to just one single quantum energy state, where all  constituents are forced to occupy. This corresponds to the zero-degree of freedom and therefore to vanishing entropy and viscosity:
\beq
\lim_{T \rightarrow 0}\eta  = \lim_{T \rightarrow 0} s  = \DD{k_B}{4\pi} \sum^\infty_{n=1} ln(\Omega_i) =\DD{k_B}{4\pi} ln(1)=0
\eeq
This implies that $\DD{\eta}{s}, $   $\eta$ should be more sensitive to temperature's variation than entropy, i.e., 
\beq
\lim_{T \rightarrow 0} ( \DD{\eta}{s}) =  \lim_{T \rightarrow 0}\DD{(\DD{d\eta}{dT})}{(\DD{d s}{dT})}
= \lim_{T \rightarrow 0}T^\gamma = 0,
\eeq
where $\gamma$ is an arbitrary positive constant. This convergence behaviour is in line with energy conservation,
as viscous energy extraction from an incompressible fluid is upper-limited by the availability of secondary energies:
$ dQ^{vis}< TdS.$
 \eit

\bit
 \item The transition from a maximally compressible into a purely incompressible fluid-phase should be associated with instantaneous input of free energy.
          This is a consequence of global energy conservation: The gravitational field, or equivalently, the energy stored in the curved spacetime
          embedding the ambient compressible media may be calculated using the surface integral  \cite[see][for further details]{Witten_EnergyTheorem1981}:
            \beq
                   \Delta E^g_{tot} = \DD{1}{16\,\pi}\int^{S^{n+1}}_{S^n}d^2 S^j (\DD{\partial}{\partial x^k}g_{jk} -  \DD{\partial}{\partial x^j}g_{kk} ),
                   \label{EnergyGrav}
           \eeq
           where the domain of integration, $\mathcal{D}^{n+1},$ is bounded by the surfaces $S^n$ and $S^{n+1},$ and $g_{jk}$ is the spacetime metric.
          As UCOs cooldown, the topology of the manifold $\mathcal{D}^{n+1}$ changes from a curved into flat one, and therefore  $\Delta E^g_{tot}$ becomes free.  As the matter has reached the maximum compressibility limit at zero-entropy,  $\Delta E^g_{tot}$ cannot be absorbed locally,  and therefore it is conjectured that this energy goes into enhancing the surface tension, i.e. the "bag energy" confining the enclosed continuum of gluon-quark medium. \\
           For $\rho = \rho^{uni}_{max},$  the energy per particle may be estimated as follows:  since the separation between baryons vanishes, i.e. $\Delta_{bb}=0$   (see Fig. \ref{BaryonIncompressiblePhase}). In this state the  distance   between quarks initially belonging to different baryons
          reduces to $\Delta_{qq}=\ell_{qq},$ which  is assumed to be the smallest possible distance between     quarks at $T=0.$
           By reducing $\Delta_{qq}$ to $\ell_{qq},$ an energy of order $ \triangle\varepsilon^+  \approx c \hbar/(2 \ell_{qq}) \approx 1/3$ GeV becomes available,  but which cannot be absorbed 
           locally\footnote{$ \triangle\varepsilon^+ $ is still lower than the energy required to raise the quarks and gluons inside
           baryons at $\rho = \rho^{uni}_{max}$ and vanishing entropy to the next excited energy state. This energy, which originates from vacuum energy between baryons,
             is termed here as "dark energy".}. $ \triangle\varepsilon^+$ may be viewed as  an excited energy state of the carriers of  the rest nuclear forces between baryons, which is subsequently converted into  tensorial surface tensions that keep the enclosed quarks and gluons
              inside super-baryons stably confined. When integrating  $ \triangle\varepsilon^+$  over all baryons that reached $\rho = \rho^{uni}_{max},$ one should  obtain an energy that is equal to the rest energy of the individual and infinitely-separated baryons. 
               For $\rho = \rho^{uni}_{max},$ this extra-energy is necessary  for facilitating the phase transition of the fermionic QGPs inside individual baryons into the  Bosonic superfluid phase, where the mass-energy of the constituents is uniformly distributed with restored Chiral symmetry. \\
              The process here is reversible, as  once    the symmetry is broken, the super-baryon starts decaying,  thereby  liberating 
              $ \triangle\varepsilon^+ =\Delta E^g_{tot}$ in a run-away hadronization procedure.\\

           A straightforward consequence of this scenario would be that the first generation of massive NSs should  have metamorphosed into invisible dark energy objects (DAOs), to end up doubling their initial mass. At a certain point in their cosmic history, the cores, would have to decay, thereby hardonizing their entire contents and giving rise to a  monstrous explosion through which each DEO would liberate  roughly $10^{57}$ GeV. Here the entire content of baryons is jettisoned into space almost with the speed of light.
           Recalling that the lifetime of neurons in free space is extremely short compared to protons, then the present scenario may nicely explain the dominant abundant of hydrogen  in our universe.\\

 \item    As DEOs are made of SuSu matter, which is characterized by a single characteristic speed: the speed of light and as they are embedded in
            purely flat spacetime, then the decaying front should propagate almost with the speed of light and would turn the object apart in less than
            $10^{-5}$ s.  The straightforward consequence here is that the ultra-relativistic outward-propagating baryons would leave the center sufficiently fast, before the embedding spacetime starts changing its curvature significantly.   At such high speeds, the baryons are too energetic and elements heavier than hydrogen
            are unlikely to form, which in turn, may  explain  the dominant abundance of hydrogen  atoms in our universe.\\
 \eit
\section{A possible origin of  $ \rho^{uni}_{cr}$}
Given  the fundamental constants, such as the Planck constant $\hbar,$ the speed of light $c$ and the gravitational constant $G,$
one may construct the following scales:
\beq
\barr{l}
\ell_p =\sqrt{ \DD{\hbar G}{c^3}} =\mathcal{ O}( 10^{-33})~cm, \\
 m_p =\sqrt{ \DD{\hbar c}{G} }= \mathcal{ O}(10^{-5})~g.
\earr
\eeq
The inclusion  of $G$ in these two expressions is purely ad hoc and supported by neither astronomical observations nor experimental data.
However, it suggests a range of scales, where the effects of both GR and quantum fields  may overlap theoretically.\\
On the other hand,  the spacetime embedding an incompressible supranuclear dense superfluid, such as inside the cores of massive NSs,
is  conformally flat \cite{Huj2020ConMap}, gravity-effects diminish, and therefore the characteristic length and time scales governing such fluids
are independent of $G.$\\
Moreover,  the incompressibility condition, i.e. $d \varepsilon/d\textrm{Vol}=0,$ implies that the separation
between the massive constituents making the mass-energy of the fluid  should saturate around a minimum length scale $\ell^{uni}_{min}$.
Recalling that the effective masses of baryons are endowed by  quarks and gluons,  $\ell^{uni}_{min}$ is expected to mainly be  related to quarks.
Indeed, the separation between nucleons in atomic nuclei, $\Delta_{bb},$ is generally larger than the radius of a single nucleon, which
can be reduced through compression exerted either by the curvature of the embedding spacetime or by momentum-exchange
during collisions with other nuclei. \\
Based on experimental and theoretical studies as well as observations of NSs, we argue that $\ell^{uni}_{min}$ should lie in the interval: \\
 \beq
        [\DD{1}{4} \leq \DD{\ell^{uni}_{min}}{ r_b} \leq 1],
 \eeq
        where $r_b$ is the radius of the a baryon at zero-temperature.   \\
 Under normal terrestrial conditions, the nuclear density is roughly $n_0 \approx  0.16\, /{\textrm{cm}}^3,$ where   the
    radius of the proton lies within the range: $[0.84 \leq r_p\leq 0.87]$  fm \cite[see][and the references therein]{Karr_ProtonRadius2019}. \\
  $\ell^{uni}_{min}< 1/4$ should be excluded as the resonance energy\footnote{i.e. maximum energy resulting from the uncertainty in the length scale: $\Delta \varepsilon_{UP} \leq c\,\hbar/ \ell$.} would largely exceed the energy required for deconfining quarks inside individual baryons. Also,  it would correspond to matter density $\rho \gg 10\times \rho_0;$ much larger than the critical density beyond
  which intermediate massive NSs  collapse into  BHs  (see Eq. \ref{BHDensity}).\\
   On the other hand,  using $\ell^{uni}_{min} \approx r_b,$ the resulting resonance energy amounts to $\Delta \varepsilon \approx 0.23$ GeV.
   Summing over all possible numbers of quark-bonds, would yield roughly the rest energy of an individual  baryon. \\
    Based thereon, the volume of a massive NS  at the end of its luminous lifetime  would be the sum of the volumes
    of the enclosed baryons at zero-temperature, i.e.
   \beq
    \mathcal{V}^{t=\infty}_{NS} = \sum^N_{n=1}{{\mathcal{V}}}^b_n = N\times {{\mathcal{V}}}^b_0,
   \eeq
   where $N, {{\mathcal{V}}}^b_0$ denote the number of merged baryons and the volume of a single baryon at zero-temperature, respectively.\\
   For an intermediate  massive NS with  $N= 10^{57}$ baryons and $\ell_{qq}=\triangle_{qq}=r_b = 0.85$ fm, we obtain a radius
    $R^{t=\infty}_{NS} = N^{1/3}r_b = 1.0168\times 10^6$ cm. While $R^{t=\infty}_{NS}$ here is  larger than the
     corresponding Schwarzschild radius, it is still smaller than the last stable orbit, though these comparisons are irrelevant,
     as the spacetime inside  $R^{t=\infty}_{NS}$ is Minkowski flat and not curved Schwarzschild.\\

 \begin{figure}[t]
\centering {\hspace*{-0.15cm}
\includegraphics*[angle=-0, width=8.0cm]{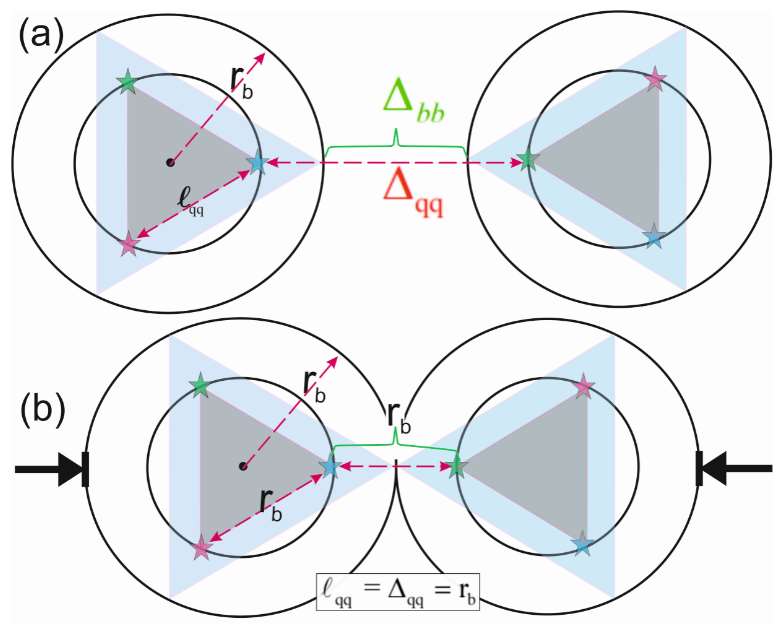}
}
\caption{\small A schematic description of  arbitrary two baryons in the maximally compressible state on the verge of merging to form
incompressible gluon-quark superfluid. Here the optimal separation between three quark flavour inside a baryon  at $\rho^{uni}_{max}$ 
and $T=0$ is $ \ell_{qq}=\triangle _{qq} =  r_b = 0.85$ fm.
}\label{BaryonIncompressiblePhase}
\end{figure}
   Recalling that the action of compression of  baryons by the embedding curved spacetime under zero entropy conditions,  is to mainly  transform the positive energy  $\Delta E^g_{tot}$ (see Eq.\ref{EnergyGrav}) into storable energy $dW.$
   From global energy conservation, one finds:
   \beq
   \Delta E^g_{tot}= dW = -Pd\mathcal{V} = -P \,(d  \sum^N_{i=1}{\Delta V}_i),
   \eeq
   where  $\mathcal{V}$ consists of the sum of  compressible space between baryons, $\Delta V_{bb},$ and those of the
   incompressible   gluon-quark plasma inside baryons at zero-temperature, ${{\mathcal{V}}}^b_0.$ For a given number N of baryons,
    we obtain  $\mathcal{V}  = \sum^N_{i=1}{ V}_i  = \sum^N_{i=1}{\Delta V}_i  + N {{\mathcal{V}}}^b_0, $ where
     ${\Delta V}_i(r)$ is a radius-dependent  quantity which is expected to increase as the surface of the object is approached.   \\
    Once the neighboring baryons are compressed together and came into direct contact with each other
    (see Fig. \ref{BaryonIncompressiblePhase}),   ${\Delta V}_i $
    reaches the lower limit:
      \beq
    {\Delta V}_i  \xrightarrow[\Delta_{qq}\rightarrow  \ell_{qq}]{} {{\mathcal{V}}}^b_0,
    \eeq
    which corresponds to the  maximum compressibility state of matter. Under these critical conditions, the available free
    energy due to spacetime compression cannot be stored locally.  One reasonable possibility would be that the quantum fluid
    undergoes a phase transition into a pure incompressible state, and the free energy goes into enhancing the pairing processes of quarks and energize the confining force of the quarks inside the super-baryon.\\
   As a consequence, the enhanced energetics of the super-baryon implies that  the embedding curved spacetime must flatten. The limiting case here would be a vanishing  gravitational energy, $\Delta E^g_{tot}.$ This energy state corresponds to the end cosmic phase of NSs, where they become invisible DEOs embedded in Minkowski spacetimes.
   In this case, the total mass-energy of a single  DEO consists  of:
   \beq
   \barr{ll}
            E_{tot}(t=\infty)&  =  E_{rest} + \Delta E^g_{tot}|_{t=0}\\
                                          &=  N  \times 0.931 \textrm{~GeV} + E_g|_{t=0}\\
                                          & = 2 \times N  \times 0.931 \textrm{~GeV},
   \earr
   \eeq
where $\Delta E^g_{tot}|_{t=0}$ is the initial gravitational energy stored in the curved spacetime embedding a newly born pulsar (see Fig.\ref{RopeProblem}).
\begin{figure}[t]
\centering {\hspace*{-0.15cm}
\includegraphics*[angle=-0, width=6.25cm]{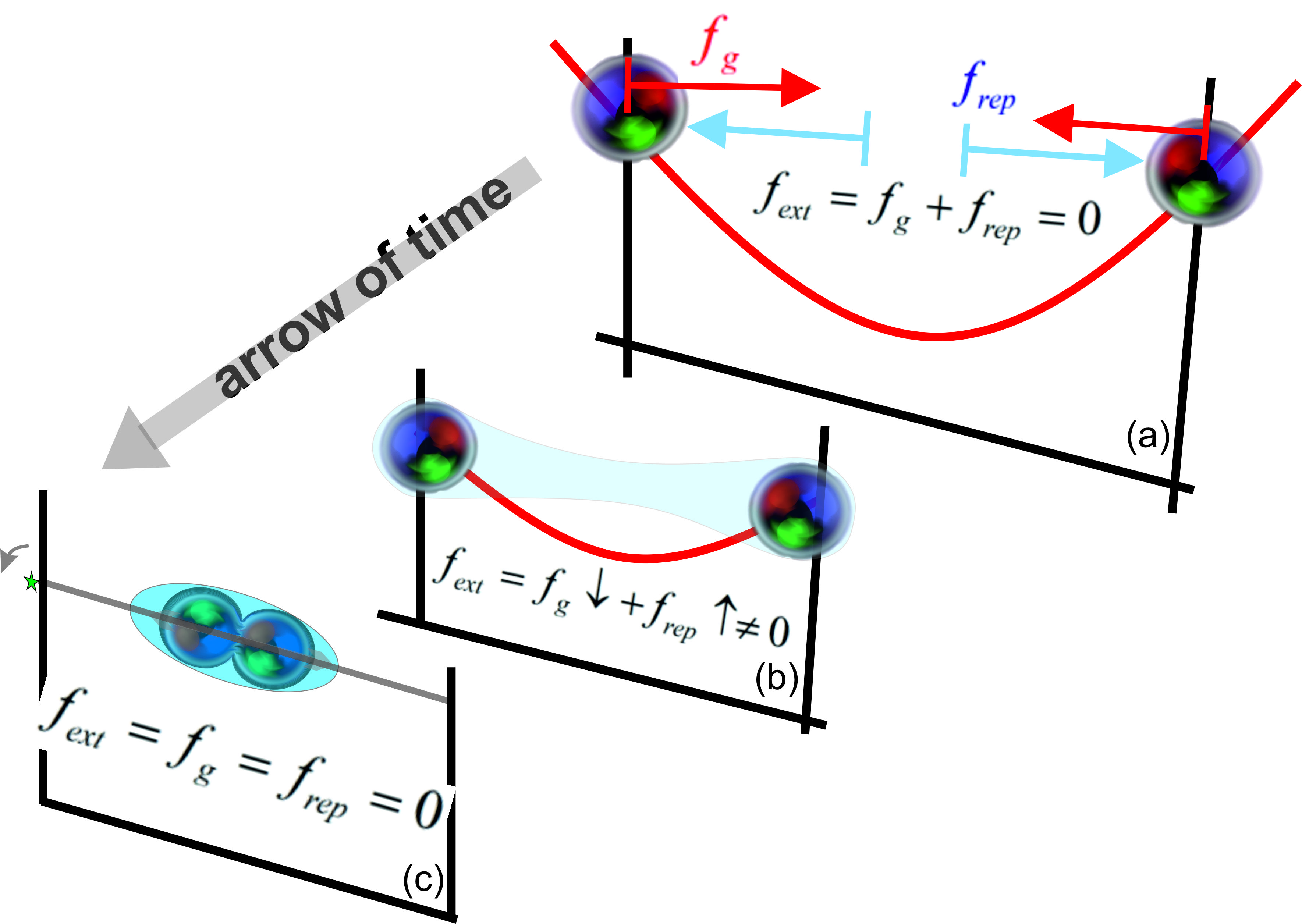}
}
\caption{\small The forces between two arbitrary baryons in hydrostatic equilibrium inside massive NSs.
The repulsive force  $f_{rep},$ is set to oppose $f_{g},$ the compressional force,  exerted by the embedding curved spacetime.
The attractive force $f_{ext},$ ist vanishingly small initially (a). As NSs spin and cool-down, the separating distance between 
baryons decreases and the strength of attractive forces, denoted by  $f_{ext},$ becomes increasingly significant (b).  The  energy associated with $f_{ext}$ is  extracted from the positive energy of the embedding curved spacetime
and therefore the spacetime should flatten. In the limit, when $\triangle_{bb} = \ell_{qq},$  baryons merge together, where the repulsive and attractive forces vanish, whilst the embedding spacetime becomes completely flat. The energy extracted from the spacetime goes into tensorial surface
tensions that confine the enclosed quarks, ensure uniform mass-energy distribution, and facilitate a phase transition into a Bosonic superfluid
with restored Chiral symmetry.
}\label{RopeProblem}
\end{figure}
  In the absence of destructive external forces, the gravitational and thermodynamical properties of DEOs suggest that these should be long-living
   objects with a lifetime most likely much longer than the current age of the universe. Once a fully evolved DEO starts decaying, its hadronization process goes instantly, thereby liberating its total $\Delta E^g_{tot}|_{t=0}.$ Recalling that the DEOs are governed by one single speed,  the speed of light, its
    hardonization  process would be  associated with a giant explosion through which the entire content of the SB is jettisoned
   into almost a flat spacetime with ultra-relativistic speeds, i.e. sufficiently fast, before the embedding spacetime starts re-curving.\\
   Due to the relatively short lifetime of neutrons in free space,  the hardonization process of the entire DEO would give rise to  creating
   hydrogen and light elements.\\
   Finally,  there are another two issues that support the present scenario: The hyperon puzzle in NS and the predicted radii  of the pre-merged NSs
   in GW170817.
   In the former case, it was shown that the inclusion of hyperons production in the EOSs  reduces the mass of the NSs significantly and make them incompatible with observations, whereas hybrid stars with cores made of deconfined quark would reverse this reduction tendency \cite[see][for further details]{Bombaci_Hyperons2016}. Also, noting that the onset of hyperon production occurs for $\rho \geq 2~ \rho_0,$ the formation of incompressible gluon-quark superfluid in their cores is a viable scenario. In the latter case, however  recent studies of the internal structure of the NSs in  GW170817  using sophisticated EOSs with  tidal deformability, predict their  radii  to be roughly $11$ km.
    Our scenario predicts that isolated NSs would end up their luminous lifetime as DEOs having radii of order $R^{t=\infty}_{NS} \approx 10$ km,
    which is a  reasonable prediction, when noting that  the volume of isolated NSs should shrink as they evolve on the cosmic time.\\
 \subsubsection{Asymptotic freedom of SuSu-matter}
  Although the density and entropy regimes of quantum fluids inside the cores of ultra-compact objects are totally different from those
  of  particle collisions under terrestrial  conditions, the QCD properties in the former case  are expected to be even simpler than in the latter case.
  Indeed,  the spatial variation of local thermodynamical quantities both in the maximally compressible and purely incompressibility fluid phases would vanish. The incompressible superfluid core is characterized by one single length scale: the radius of the core, $R_{core},$ and
   one single time scale $\tau_{core}= R_{core}/c.$ As all constituents occupy the same quantum state, the momentum transfer $Q^2$ between
   them  saturates around the maximum possible value, where the divergence of the tensorial momentum,
   $\mathbf{Q}^{\mu\nu}$ are forbidden, i.e.
   \[   \textbf{Q}^{\mu\nu}_{;\nu} = 0
   \]
                  In terms of the  renormalization  group equation for the running coupling constant  $\alpha_s$ \cite[see][and the references therein]{Gracey_AsymptoticFreedom2020}:
                  \beq
                   Q^2 \DD{\partial \alpha_s(Q^2)}{\partial Q^2}  = \beta (\alpha_s(Q^2)),
                  \eeq
                  the conditions to be imposed on $ Q^2$ at both boundaries are identical and therefore $  \alpha_s(Q^2)$ must remain constant.
                  In fact the Chiral symmetry is fully restored $(\langle \bar{q}q\rangle =0)$   in this regime, which grants the core dynamical stability
                  on cosmic timescales.

\subsection{The bimetric universe and CMB radiation}
While the existence of a universal maximum energy density promotes the hypothesis that our universe
is infinitely large and old,  it suggest that it went neither through an inflationary phase nor creating BHs. The model is still conforms with the smooth and uniform
temperature of the cosmic microwave background radiation (CMB). Here the instant decay of a finite and perfectly
symmetric cluster of  DEOs, whose contents are made of quark-gluon superfluids with restored Chiral symmetry
may have the potential of creating the CMB we observe today. In this case, our location in the embedding spacetime
must have been pretty close to the center of a spherically symmetric gigantic explosion, which is termed here as a
 local big bang (LBB). The progenitor of the LBB is a giant cluster of DEOs embedded in  flat spacetimes.
Have the cluster started decaying, its whole content undergo a rapid hadronization process, and the energy
stored  in the system undergo a reversible process: part of this energy goes back into curving the embedding spacetime,
while the rest goes into forming a soup of highly energetic particles and radiation.\\
Here a race starts between the outward-oriented ultra-relativistically propagation of  fluid flows and topology fitting
of the embedding spacetime.  Hence two fronts were created: The hardonization front that must go through the entire cluster, and the
topological front, $``f_1$'' at the interface between the flat the curved spacetimes (see Fig.\ref{UniverseBBangQSOs}). Although the material front
propagates at ultra-relativistic speed,  the topological front $``f_1$'' propagates with the speed of light. This generates a time delay that increases with the cosmic age.
Note that the hardonization front is associated with the creation of interacting particles, generating entropy, and turning
the matter into compressible dissipative media. In the course of this process, the embedding spacetime ought to undergo a topological change from Minkowski flat into a curved Schwarzschild. However, as the amount of energy that goes into curving the embedding spacetime is finite,
the curvature should decrease as the expansion goes on, thereby asymptotically converging into a perfectly flat spacetime (see Fig. \ref{UniverseBBangQSOs}).   \\
Mathematically, let the rest mass-energy of the cluster of DEOs on the verge of an LBB explosion  be $\mathcal{E}_{cl}.$  As the
cluster is embedded in a flat spacetime, the gravitational energy vanishes, i.e.. $ \mathcal{E}_g = 0.$  Assuming mass-energy conservation, then
the total energy shortly before and after LBB should remain constant, i.e.:
\beq
\{[\mathcal{E}^b_{cl} + \mathcal{E}^{vac}_{cl}]  + \mathcal{E}_g\}_{t\leq 0}  =
\{\mathcal{E}^b_{cl} +[\mathcal{E}^{vac}_{cl}  + \mathcal{E}_g]\}_{t> 0},
\eeq
 where $\mathcal{E}^b_{cl}$ is the rest energy of  the total baryons at the nuclear density, $\rho_0,$ and zero-entropy. $\mathcal{E}^{vac}_{cl}$ is the  work needed for compressing  the constituents in the system from  $\rho_0,$ up to the maximum compressibility limit, where
 $\rho=\rho^{uni}_{max}.$  $ \mathcal{E}_g$   is the gravitational energy, which is  calculated from the integral:
 \beq
 \barr{lll}
  \mathcal{E}_g & =&  \DD{1}{16\,\pi}\int^{S^{n+1}}_{S^n}d^2 S^j (\DD{\partial}{\partial x^k}g_{jk} -  \DD{\partial}{\partial x^j}g_{kk} ) \\
                          & =& \left\{
                                        \barr{lll}
                                                            0                      &   & \textrm{flat spacetime}\\
                                                  \mathcal{E}^b_{cl}  &  &  \textrm{Schwarzschild spacetime}.
                                          \earr
                                          \right.
  \earr
\eeq
 Based thereon, the total mass involved in the big bang may be predicated as follows: for an average density of $\rho_{now}\approx 10^{-29}$ g/cc, and
 a radius of $R^{uni}> c\times \tau^{uni}_{age} \approx 10^{28} $ cm, we obtain a total mass of  $\mathcal{E}^b_{cl}\approx 5\times 10^{22}~ \MSun.$  Inside the DEO cluster the density is roughly uniform and has the value $\rho^{uni}_{max}.$ This yields
  a radius of approximately $2~AU$ for the DEO cluster. Note that a SMBH with $R_S = 2~AU$ yields an enclosed mass of roughly
  $10^8~\MSun$ which falls in the  lower  mass-regime of currently observed quasars. Alternatively, a SMBH of  $10^{22}\MSun$
  would yields  Schwarzschild radius of order $10^{27},$ which should be ruled as well. \\
  Indeed, the cluster of DEOs presented here  has a uniform density and is embedded in a Minkowski spacetime, so that the necessity
  for an exponential growth to smooth out density irregularities becomes it unnecessary.

 \begin{figure}[t]
\centering {\hspace*{-0.15cm}
\includegraphics*[angle=-0, width=8.0cm]{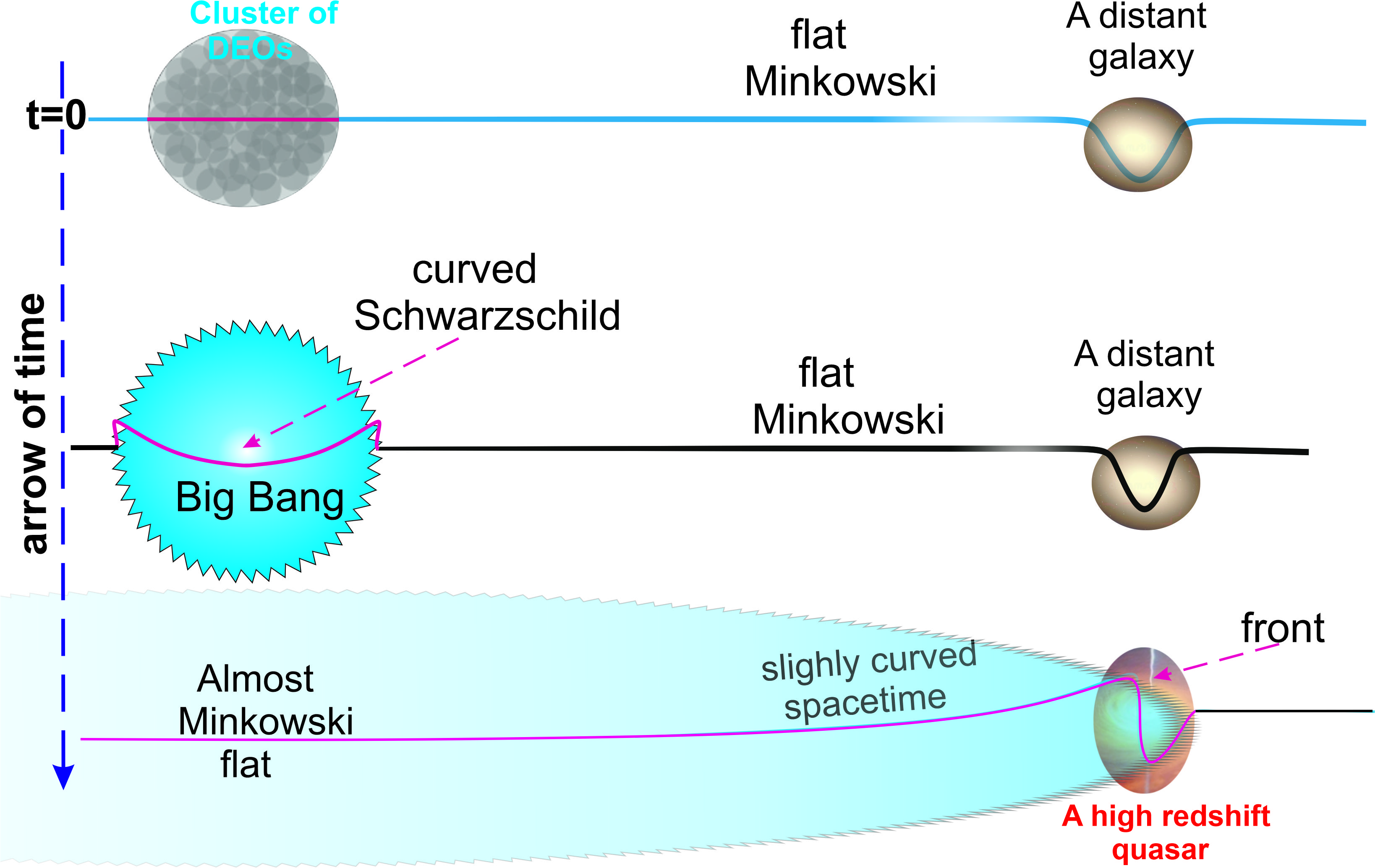}
}
\caption{\small The structure of the bimetric spacetime of our expanding universe. Accordingly, the progenitor of the big bang is a cluster of DEOs that conglomerated over dozens of Gyr at the background of an infinitely flat spacetime.
Immediately after the instantaneous  decay of the cluster into normal dissipative matter, the topology of  the local flat spacetime changed into
a curved Schwarzschild, thereby initiating a spherically symmetric expansion front, that  separated the two topologies and propagated with the speed of light. Had the expansion front marched through distant, quiet and old galaxies that are embedded in locally curved spacetimes, but superimposed
on an infinitely flat spacetime,  then the motions of their entire contents of objects would be considerably perturbed and turn turbulent, thereby
 setting these galaxies into active mode, that identified today as powerful high redshift QSOs.
}\label{UniverseBBangQSOs}
\end{figure}

.
\subsection{Summary \& Discussion}
In this paper, we have discussed the possibility that our universe may permit the existence of a universal maximum energy density $\rho^{uni}_{max},$ which characterizes the state of  incompressible superfluids at the center of massive NSs.
  Based on theoretical and observational studies of pulsars
and neutron stars, $\rho^{uni}_{max}$ is predicted to lie between  $2\times \rho_0$ and  $3\times\rho_0. $  Under these conditions
 and at zero-temperature,  the irreducible distance between quarks is predicated to be of order:  $\ell_{qq}\approx 0.85$ fm, i.e. to the experimentally revealed value of  the radius of a stable baryon.\\
 The underlying assumption here is that UCOs should be embedded in bimetric spacetimes: their incompressible
 quark-gluon superfluid cores are embedded  in Minkowski-type spacetimes, whereas the ambient compressible and dissipative
 media  are embedded in curved spacetimes.\\
 Based on this bimetric spacetime scenario and on the universal maximum density hypothesis,  solutions to several debates both in physics and
 astrophysics could now be provided:\\
 \begin{itemize}
   \item Incompressible fluids are treatable using the field equations of General Relativity and may be modeled in a self-consistent manner,
   without violating the causality condition or creating physically unrealistic ultrabaric regimes \cite[see Fig. 7 in ][]{Hujeiratmetamorph2018}.
   \item The quantum mechanisms underlying the glitch phenomena in pulsars, the increasing intervals between successive glitches as the objects age,  the origin of under and over-shootings  of the rotational frequencies observed to associate the glitch events of the Vela pulsar
       can be well-explained.
   \item Formation of stable massive NSs without collapsing into stellar BHs, such as the remnant of GW170817, may be considered as a reliable  evolutionary track.  In this case the remnant of  the first generation of massive NSs should have fully metamorphosed into invisible DEOs
       by now.
\item The cosmological origin of high redshift QSOs  and  dark matter in the early universe conform with our scenario that the universe should be infinitely large and old.
\item The dominance and origin of chemical abundance of the light elements in the universe is a natural consequence of  the collective decay
      clusters of DEOs  that led to LBB explosion through which hardonizion operated efficiently.
\item The scenario of a bimetric universe provides a simple and reasonable explanation to the flatness problem of the universe, the origin of dark matter and dark energy.
\item The present scenario is capable of providing a new robust mechanism for the origin and power of high redshift QSOs.
         Indeed, the topological front that separating the Schwarzschild from the Minkowski spacetimes carries sufficient energy
         to significantly deform the locally curved spacetime embedding distant galaxies, thereby turning the well-ordered motions of
         their contents of objects into turbulent motion. This excited energy state would considerably increase the rate of destruction and collision
         of stars and would turn the galaxy into an active mode for a considerable cosmic time.\\
         The duration of this activity starts from the moment at which the topological front collides with locally curved spacetime embedding
         the respective galaxy and would continue until the  remaining stellar components have settled back into a significantly lower energy state
         \cite[see][for further details]{Hujeirat_Redshift_QSO2021}.
 \end{itemize}

 One of the far-reaching consequences here is that the big bang should be a recurrent phenomenon in our infinitely large and old universe.
Indeed,  the existence of a maximum universal density $\rho^{Uni}_{cr}$ would render the inflationary phase unnecessary and
would naturally explain why our universe escaped its collapse into a  giant BH.\\
 The scenario here predicts that massive stars should have sufficient time to collapse into massive UCOs, cooldown and to subsequently
 turn invisible at the end of their luminous lifetimes.
Their further collapse into BHs is prohibited by the existence of  $\rho^{Uni}_{cr},$ at which the topology of the embedding spacetime
changes into a Minkowski-type spacetime (see Hujeirat 2021 in preparation).\\
Nevertheless, our scenario addresses several new problems that need to be answered, such as:
\begin{enumerate}
  \item What is the physical nature of the well-observed stellar and massive black holes?
  \item What are the physical mechanisms that drive DEOs to conglomerate into clusters, how DEOs interact with each other
  or with other gravitationally bound astrophysical objects?
  \item What are the predicated lifetimes of DEOs, and what are the mechanisms underlying their decay?
\end{enumerate}

 \end{document}